\documentclass[prl,aps,twocolumn]{revtex4-1}
\usepackage{bm}
\usepackage{graphicx}
\usepackage{amsmath}
\usepackage{amssymb} 
\usepackage[utf8]{inputenc}
 \usepackage[T1]{fontenc}
\usepackage{color}
\usepackage{upgreek}
\usepackage{subfigure}
\usepackage[unicode=true,colorlinks=true,citecolor=blue]{hyperref}
\usepackage{placeins}
\usepackage{lipsum}
\usepackage{epsfig}
\usepackage[utf8]{inputenc}
\usepackage{wrapfig}
\usepackage[normalem]{ulem}
\newcommand{\nix}[1]{}
\newcommand{\sign}{\mathop{\rm sign}}

\newcommand{\pderiv}[2]{\frac{\partial #1}{\partial #2}}
\renewcommand{\phi}{\varphi}

\renewcommand{\i}{\mathrm i}

\newcommand{\eps}{\varepsilon}

\newcommand{\beq}{\begin{equation}}
\newcommand{\eeq}{\end{equation}}
\newcommand\beqa{\begin{eqnarray}}
\newcommand\eeqa{\end{eqnarray}}
\newcommand\ba{\begin{array}}
\newcommand\ea{\end{array}}

\renewcommand\d{\partial}

\newcommand\nn{\nonumber}

\begin{document}

\title{Edge photocurrent driven by THz electric field in bi-layer graphene}

\author{S. Candussio$^1$,  M.V. Durnev$^2$, S.A. Tarasenko$^2$, J. Yin$^3$, J. Keil$^1$, Y. Yang$^3$, S.-K. Son$^3$, A. Mishchenko$^3$, H. Plank$^1$, V.V. Bel'kov$^2$, S. Slizovskiy$^{3,4,6}$, V. Fal'ko$^{3,4,5}$,  and S.D. Ganichev$^1$
}

\affiliation{$^1$Terahertz Center, University of Regensburg, 93040 Regensburg, Germany}
\affiliation{$^2$Ioffe Institute, 194021	St. Petersburg, Russia}
\affiliation{$^3$Department of Physics \& Astronomy, University of Manchester, Manchester M13 9PL, UK }
\affiliation{$^4$National Graphene Institute, University of Manchester, Manchester M13 9PL, UK }
\affiliation{$^5$Henry Royce Institute for Advanced Materials, Manchester, M13 9PL, UK}
\affiliation{$^6$St. Petersburg INP, Gatchina, 188300, St.Petersburg, Russia}

\begin{abstract}
 We report on the observation of edge electric currents excited in bi-layer graphene by terahertz laser radiation. We show that the current generation belongs to the class of second order in electric field phenomena and is controlled by the orientation of the THz electric field polarization plane. Additionally, applying a small magnetic field normal to the graphene plane leads to a phase shift in the polarization dependence. With increasing the magnetic field strength, the current starts to exhibit $1/B$-magnetooscillations with a period consistent with that of the Shubnikov-de Haas effect and amplitude by an order of magnitude larger as compared to the current at zero magnetic field measured under the same conditions. The microscopic theory developed shows that the current is formed in the edges vicinity limited by the mean-free path of carriers and the screening length of the high-frequency electric field. The current originates from the alignment of the free carrier momenta and dynamic accumulation of charge at the edges, where the $P$-symmetry is naturally broken. The observed magnetooscillations of the photocurrent are attributed to the formation of Landau levels.
\end{abstract}

\maketitle

\section{Introduction}
\label{introduction}

Bi-layer graphene (BLG) with the band gap and conductivity tunable by a gate electric field is in the focus of state-of-the-art carbon electronics and optoelectronics \cite{Novoselov2006, McCann2006, Ohta2006, Castro2007, Abergel2007, Geim2007, Morozov2008, Neto2009, Dean2010, McCann2013}. Of particular interest are second-order nonlinear phenomena such as second harmonic generation, photogalvanic, photon drag, plasmonic, photothermoelectric, and ratchet effects, where the $ac$ electric field of the laser radiation drives an $ac$ electric current at the double frequency or a $dc$ electric current~\cite{SturmanFridkin1992,ivchenko1997,Falko1989,Tarasenko2011,Budkin2016,GlazovGanichev_review,Koppens2014,Quereda2018,Koppens2019}. In bi-layer graphene so far only bulk mechanisms of the current generation were considered~\cite{Kumar2013, Dean2009, Peres2014, Wu2012, Lin2014,Spirito2014,Brun2015,Kheirabadi2016,Kheirabadi2018, Avetissian2019, Hipolito2016, Yang2014, McGouran2016,Bandurin2018}. Here, we show that $P$-symmetry breaking, required for the second-order effects to occur~\cite{SturmanFridkin1992}, is brought by the sample edges~\cite{Karch2011} and that the effect is observed in high-mobility $\mu$m-scale hBN/BLG/hBN structures excited by an $ac$ electric field of polarized terahertz laser radiation. The photocurrent flows along the sample edges and its magnitude and direction are controlled by the relative orientation  of the electric field polarization vector and the corresponding edge. Variation of the back gate voltage reveals that the photocurrent directions are opposite for $p$- and $n$- conductivities and the photocurrent magnitude is a non-monotonic function of the gate voltage. 

\begin{figure}
	\centering
	\includegraphics[width=\linewidth]{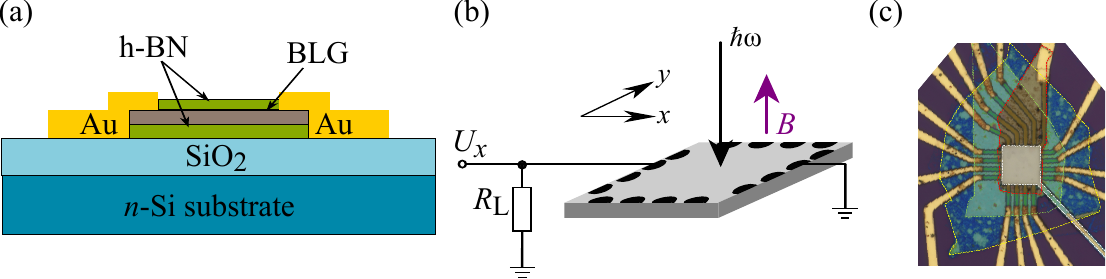}
	\caption{ (a) Sketch of the structure cross section. (b) Schematic of the experimental set up shown for sample \#A. Experimental arrangement is shown for  normal incident THz radiation and an external magnetic field $\bm B$ applied perpendicularly  to the graphene layer. (c) Optical micrograph of sample \#A.}
	\label{fig1}
\end{figure}

\begin{figure}
	\centering
	\includegraphics[width=\linewidth]{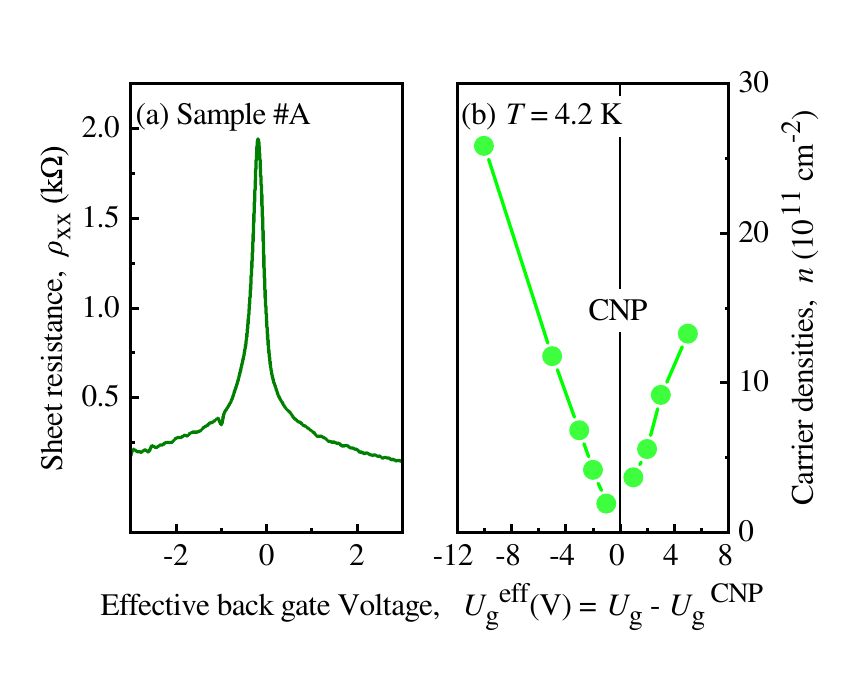}
	\caption{ (a) Longitudinal resistance as a function of the effective back gate voltage $U_g^{\mathrm{eff}}=U_g - U_g^{\mathrm{CNP}}$. (b) Gate voltage dependence of electron (positive $U_g^{\mathrm{eff}}$) and hole (negative $U_g^{\mathrm{eff}}$) densities. 
	}
	\label{fig2}
\end{figure}

Within this study we develop a microscopic theory of such an edge photogalvanic effect where free carriers are driven by the $ac$ electric field of the terahertz radiation. The photocurrent  is formed in a narrow channel with the width determined by the carrier mean free path and the screening length of the high-frequency electric field at the edge and consists of two contributions. The first contribution originates from the alignment of the free carrier momenta by the high-frequency electric field. Photocurrents associated with the alignment of the electron momenta and related phenomena were previously studied at the surfaces of bulk semiconductor crystals~\cite{Magarill1979,Alperovich1981,Alperovich1989,Schmidt2016}, metal films~\cite{Magarill1981,Gurevich1993,Mikheev2018}, and recently in graphene~\cite{Karch2011,Golub2011,Hartmann2011}.  The second contribution can be interpreted as the interplay of the dynamical charge accumulation near the edge~\cite{Volkov88,FalkoIordanskii,Kukushkin04} and the oscillating carrier density driven by the $ac$ electric field resulting in a  $dc$ current flowing along the edge. 

In the THz regime, where $\omega \tau > 1$,  the edge current $J_y \propto (E_x E^*_y + E_y E^*_x)$ is the largest for the radiation polarized at $\pm \,\pi/4$ to the sample edge. Here, $\omega$ is the angular radiation frequency, $\tau$ is the momentum relaxation time, and $E_x$ and $E_y$ are  the amplitudes of $ac$ electric fields perpendicular to and along the edge, respectively.  Turning on the magnetic field $\bm B$ normal to the BLG plane results in the rotation of the electron distribution in the momentum space shifting thereby the optimal polarizations of the THz radiation to the angles $ \alpha_{\rm max} (B) =\pm( \pi/4 + \theta_B/2)$ with respect to the sample edge, where $\theta_B = \arctan (2 \omega_c \tau)$, and $\omega_c$ is the cyclotron frequency. Upon increasing the magnetic field, the angle $\theta_B \to \pi/2$, with the maximal edge current generated by the $ac$ field polarized perpendicular to or along the edge. At strong magnetic fields and low temperatures, the photoresponse develops $1/B$  magnetooscillations correlated with the Shubnikov-de Haas (SdH) oscillations. The amplitude of the magnetooscillations in the photoresponse is an order of magnitude larger than the photoresponse at zero magnetic field measured under the same conditions.

\section{Samples and methods}
\label{samples_methods}

Experiments were carried out on bi-layer graphene encapsulated in hexagonal boron nitride. The structure cross section is shown in Fig.~\ref{fig1}~(a). Two different types of samples were studied. The first (sample \#A) was fabricated in van der Pauw geometry with four gold contacts on each edge, see Fig.~\ref{fig1}~(b). The second (sample \#B) is a Hall bar structure. Both samples were equipped with back gates.  Sweeping the back gate voltage the charge neutrality point (CNP) is well observable in the longitudinal resistance, see Fig.~\ref{fig2}~(a). The CNP slightly shifts between different cool down cycles, therefore, we introduce an effective gate voltage $U_g^{\mathrm{eff}}=U_g - U_g^{\mathrm{CNP}}$. Figure~\ref{fig2}~(b) shows the gate voltage dependence of the densities for electrons/holes.   

For optical excitation we used two different types of THz lasers. The first system was a continuous wave (\textit{cw}) molecular gas laser optically pumped by a \textit{cw} CO$_2$ laser~\cite{Ganichev2009,Olbrich2013}. In the described experiments we applied  radiation at frequencies $f= 2.54$ and $1.62$~THz, which were obtained with methanol and difluormethane as active media, respectively. The radiation power on the samples was about 40~mW. In order to modulate the radiation a chopper at a frequency of about $130$~Hz was placed in front of the laser. As second system a pulsed molecular laser was used, operating at frequency $f=1.1$~THz~\cite{Ganichev1998,Ganichevbook,Weber2008}. Single pulses with a duration in the order of 100~ns and peak  power in the order of tens of kilowatts were used. The pulsed laser was pumped by a transversely excited atmospheric pressure (TEA) CO$_2$ laser. The beam cross-section had the Gaussian shape, which was monitored by a pyroelectric camera~\cite{Ziemann2000}. The spot sizes at the full width at half maximum are about 2~mm. To control the incidence power of the laser terahertz pyroelectric and photon drag detectors were used. To modify the polarization state of the laser radiation, initially linearly polarized along the $x$-axis, we used  crystal quartz $\lambda/2$-plates. By that the orientation of the radiation electric field vector was defined by the azimuthal angle $\alpha$. 

 Magnetic field up to 4~T and the THz radiation were applied normal to the bi-graphene layer. Generated photocurrents $J$ were measured as a voltage drop $U$ over a load resistor of $R_L = 10~\mathrm{M} \Omega$ or $R_L = 470~\Omega$.  A lock-in amplifier  or a digital oscilloscope were used for recording the photoresponses for measurements with cw ($J\propto U/R_s$) and pulsed ($J\propto U/R_L$) laser systems, respectively. Here $R_s$ is the sample resistance measured between contact pairs used for measurement. The corresponding measurement setup is shown in Fig.~\ref{fig1}~(b). 

\section{Results}
\label{results}

\begin{figure}
	\centering
	\includegraphics[width=\linewidth]{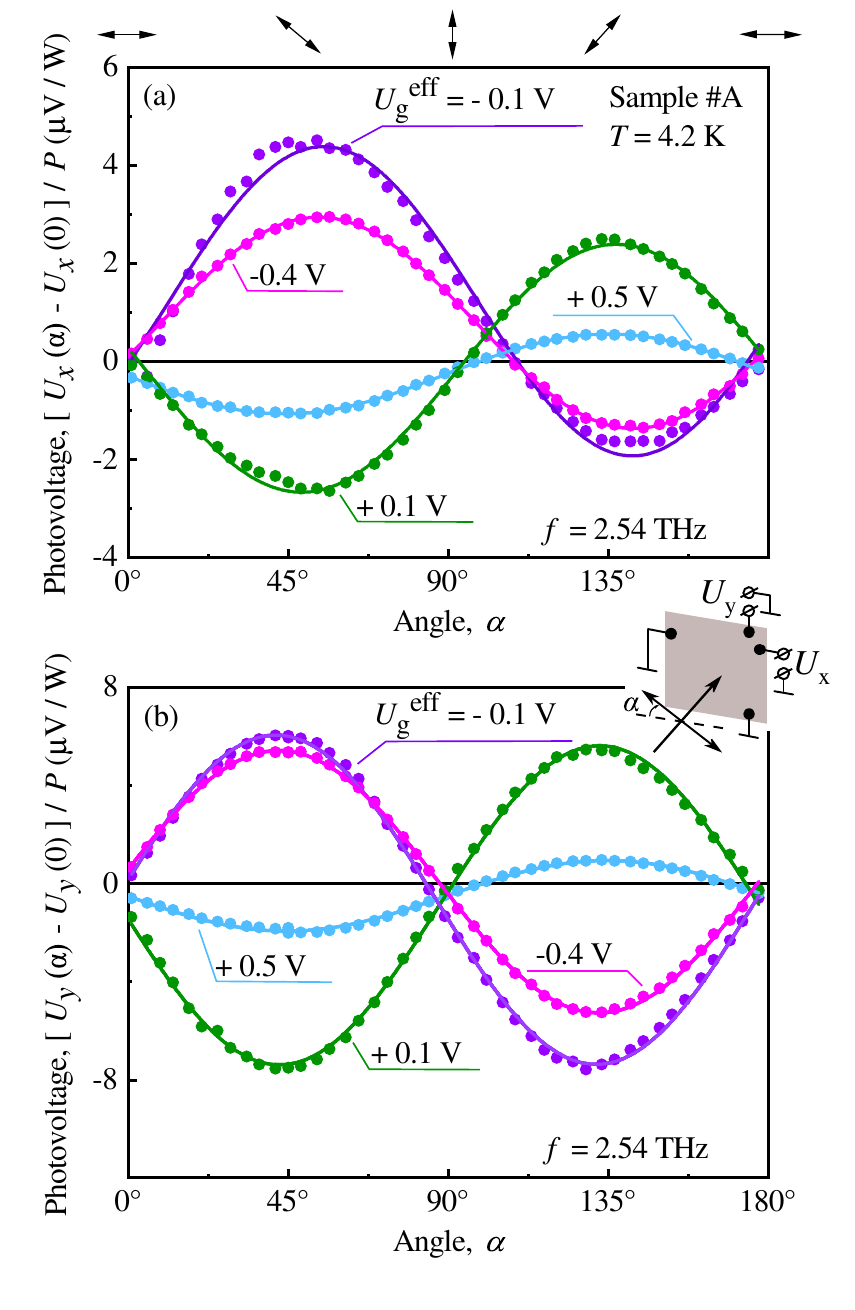}
	\caption{ Dependencies of the normalized  photovoltage $U_{y,x}/P\propto J_{y,x}/P$ on the azimuthal angle $\alpha$ obtained for contact pairs along top edge, panel (a), and right edge, panel (b),  see the inset. Note that for better visualization a small 	 signal at $\alpha =0$ is subtracted. The data are obtained for radiation frequency 2.54~THz and presented for several values of the effective gate voltage $U_g^{\rm eff}$. Arrows on the top of panel (a) illustrate the orientations of the radiation electric field vector for several values of $\alpha$.
	} 
	\label{fig3}
\end{figure}

\begin{figure}
	\centering
	\includegraphics[width=\linewidth]{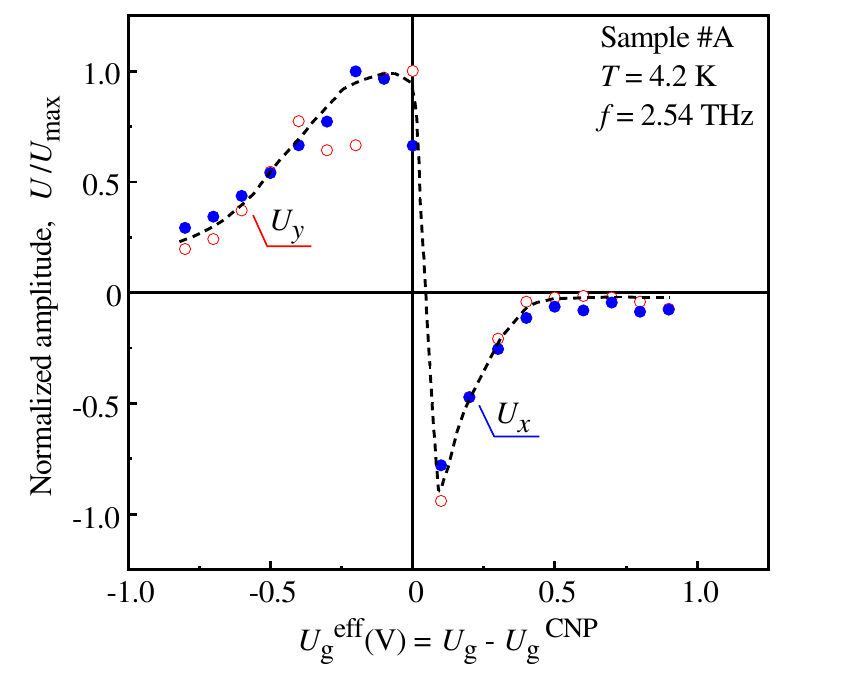}
	\caption{ Dependence of the photovoltage amplitude $U^L_{y,x}\propto J^L_{y,x}$ on the  effective gate voltage $U_g^{\rm eff}$. 
	}
	\label{fig4}
\end{figure}

First we discuss the results obtained for zero magnetic field. A photosignal, picked up from any contact pairs along one of the samples edges was detected applying normally incident linearly polarized radiation. The magnitude and the direction of the observed photocurrent depend on the relative orientation of the radiation polarization plane and the edge along it is measured. Figure~\ref{fig3} shows  polarization dependencies of the edge photosignals $U_{x,y} \propto J_{x,y}$ measured for  two perpendicular edges of the square shaped graphene sample. The data can be well fitted by 
\begin{eqnarray} \label{N1}
\label{helicity_eqn}
U_{x,y} \propto J_{x,y} =J^{L}_{x,y} \sin (2\alpha +\varphi_0) +J^0_{x,y}\:,
\end{eqnarray} 
with the azimuthal angle $\alpha$,  the  photocurrent amplitudes $J^L_{x,y} \propto U^L_{x,y}$ and a small phase shift $\varphi_0$. Note that a small polarization independent offset with $U^0_{x,y} \ll  U^L_{x,y}$,  was detected for some conditions. It may be caused by photothermoelectric effects, see e.g. Ref.~\cite{Koppens2019}, and its origin is out of scope of the present paper. Comparison of the signals obtained for neighbouring edges at fixed orientation of the polarization plane, e.g. $\alpha =45^\circ$ or $135^\circ$, shows that the photocurrents flow either towards the samples corner formed by these edges or away from it.  For instance, for negative gate voltages, for $\alpha =45^\circ$ we obtained $J_x >0$ and  $J_y>0$ whereas for $\alpha =135^\circ$ the currents become $J_x <0$ and $J_ y < 0$, see Fig.~\ref{fig3}. Measurements at different  gate voltages show also that the signal changes its sign in the vicinity of the charge neutrality point, i.e. by changing type of carriers from holes to electrons, see Figs.~\ref{fig3} and \ref{fig4}. 

The observed polarization dependence of the response reveals that it is caused by edge photocurrents. Indeed, only in this case the both signals measured along two orthogonal edges are characterized by Eq.~(\ref{N1}). In general, a bulk photogalvanic current can also be generated in bi-layer graphene with structure inversion asymmetry which has no center of space inversion and is described by the C$_{3v}$ point group. For systems belonging to this symmetry, two orthogonal components of the bulk photocurrent excited by normally incident radiation are proportional to $\cos 2\alpha$ and $\sin 2\alpha$ ~\cite{Olbrich2014}. As a results, the $\pi/2$ phase shift would be present between the photocurrents $J_x$ and $J_y$. In our experiments, however, the photocurrents flowing in the orthogonal directions are described by identical polarization dependencies. This fact clearly demonstrates the  edge-related mechanism of the detected photoresponse.

\begin{figure}
	\centering
	\includegraphics[width=\linewidth]{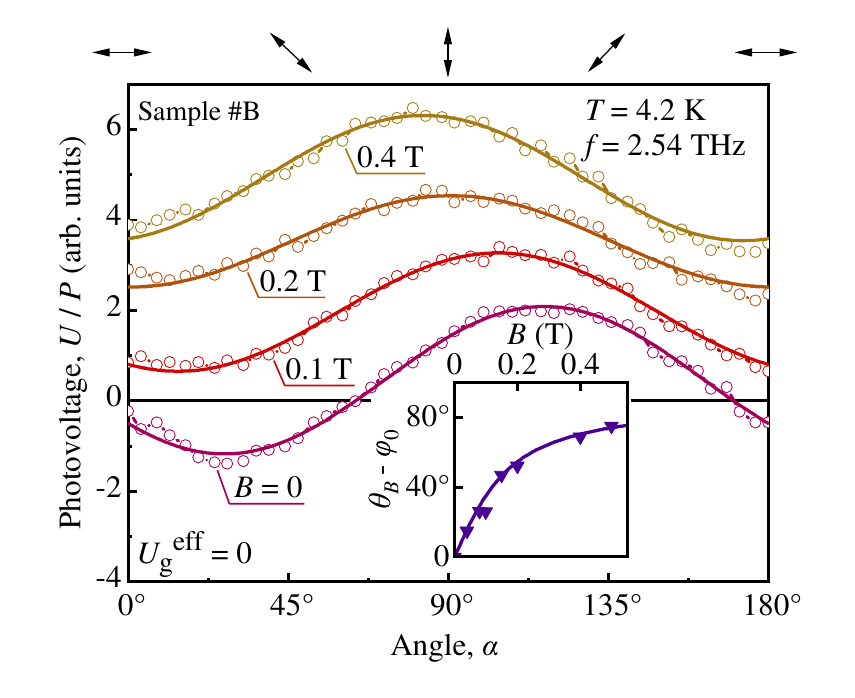}
	\caption{ Dependencies of the normalized  photovoltage $U/P\propto J/P$ on the azimuthal angle $\alpha$ obtained for contact pairs along the edge of Hall bar sample \#B.  The data are obtained in the vicinity of the CNP for a radiation frequency of 2.54~THz and presented for several values of the external magnetic field $B$. Curves are fits after $U = U^{L}(B) \sin ( 2\alpha + \varphi_0 + \theta_B)$. Note that for better visualization curves and data are vertically shifted. Arrows on top illustrate the orientations of the radiation electric field for several values of $\alpha$. The inset shows the measured phase shift $\theta_B$ (triangles) and its model prediction $\theta_B = \arctan (2 \omega_c \tau)$ calculated for $\tau = 0.6$~ps (line). } 
	\label{fig5}
\end{figure}

\begin{figure}
	\centering
	\includegraphics[width=\linewidth]{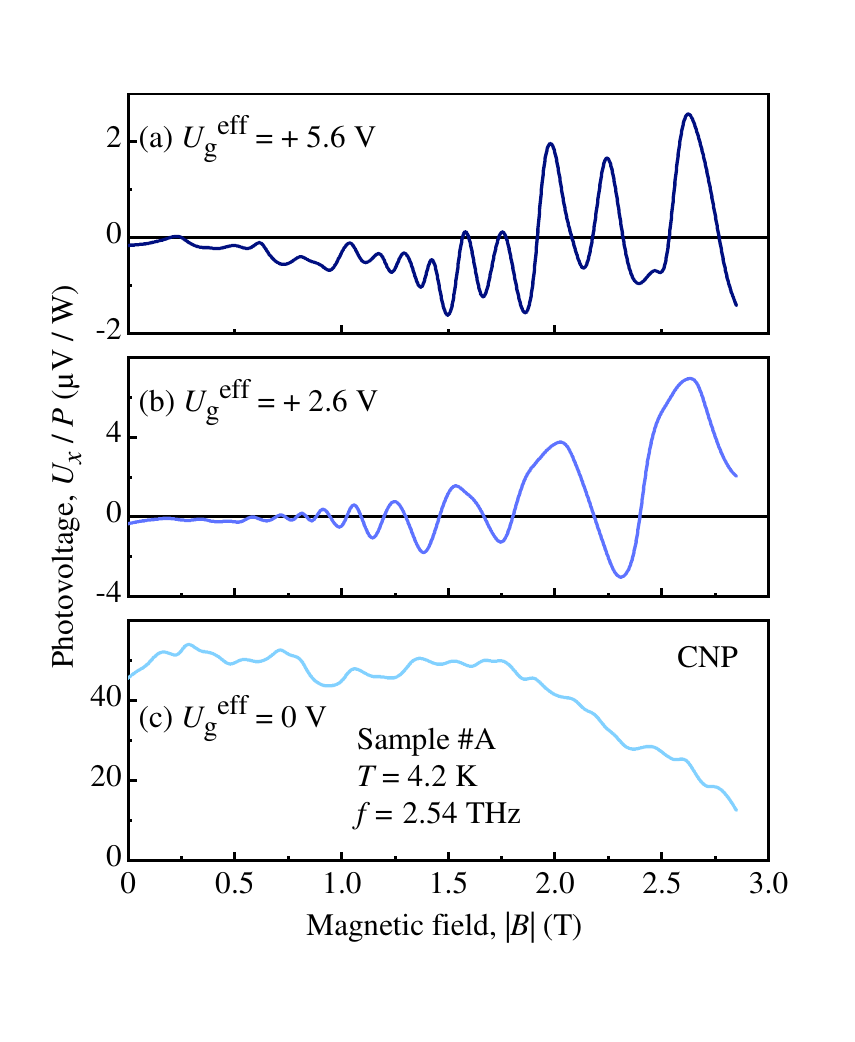}
	\caption{ Magnetic field dependence of the normalized photoresponse $U_x/P$ measured along  top edge for different effective gate voltages. }
	\label{fig6}
\end{figure}

\begin{figure}
	\centering
	\includegraphics[]{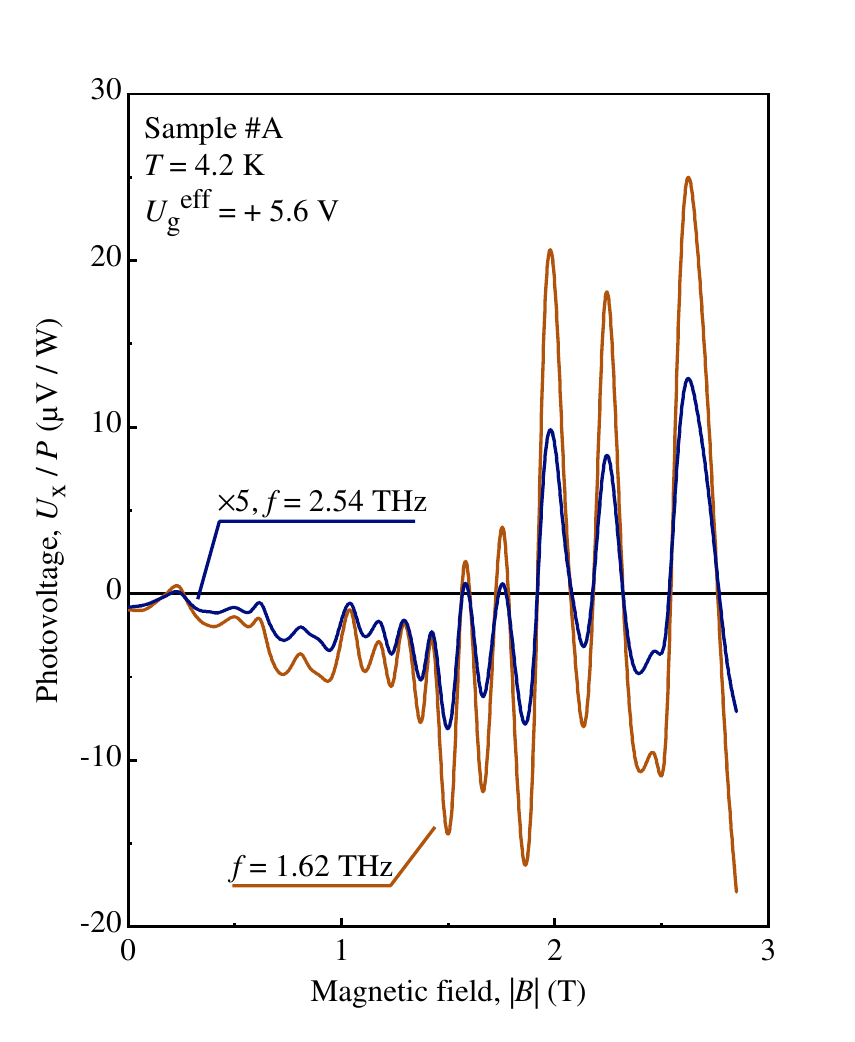}
	\caption{ Magnetic field dependencies of the normalized photoresponse $U_x/P$ measured along the top edge for two radiation frequencies. Note that the data for $f=2.54$~THz are multiplied by factor 5.  }
	\label{fig7}
\end{figure}

\begin{figure}
	\centering
	\includegraphics[width=\linewidth]{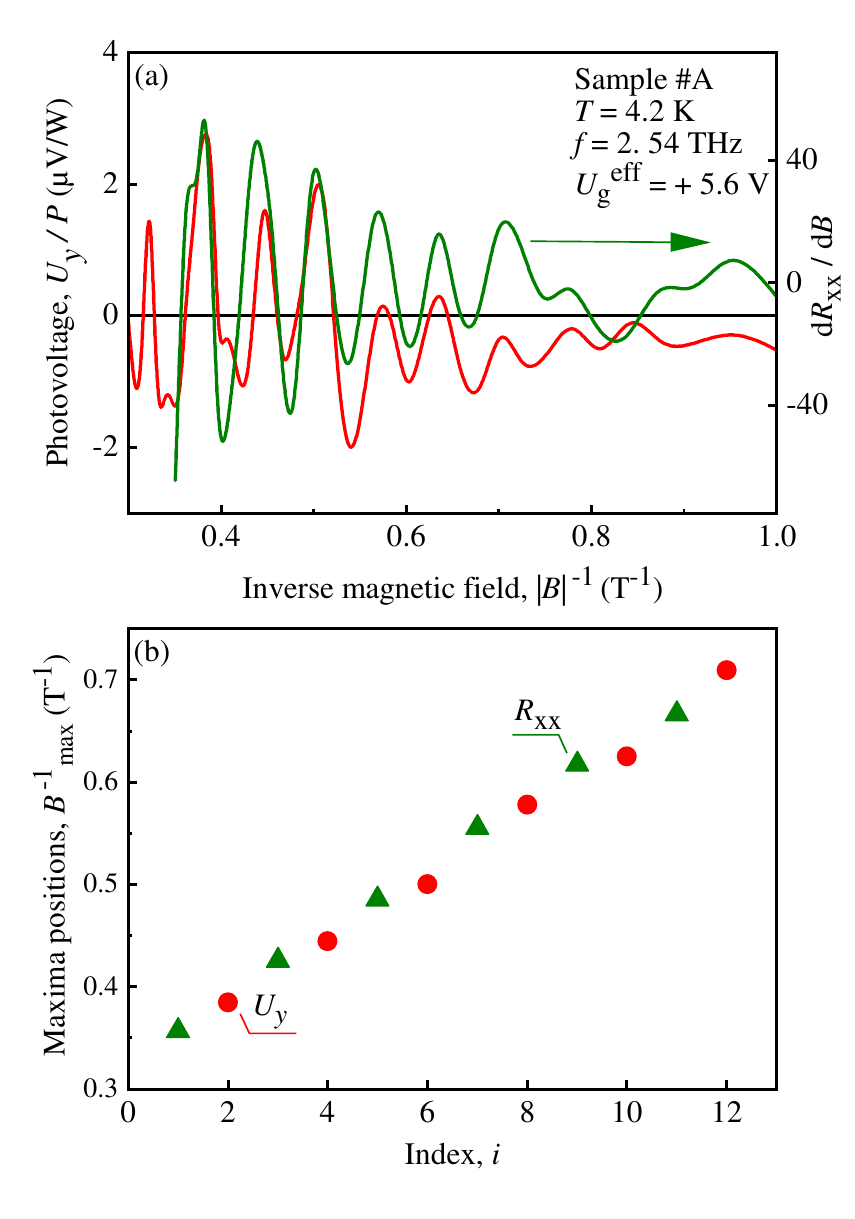}
	\caption{ (a) Magnetic field dependence of the normalized photoresponse $U_x/P$ measured along the right edge  (red curve) and of the first derivative of the longitudinal resistance (green curve). (b) Position of maxima of the photoresponse (red dots) and magnetoresistance (green triangles) are plotted against the oscillation numbers. }
	\label{fig8}
\end{figure}

At low magnetic fields the polarization dependence changes to $U_{x} = U^{L}_{x}(B) \sin ( 2\alpha + \varphi_0 + \theta_B) $, and for magnetic field of about 0.4~T we detected an almost $45$ degree phase shift, see Fig.~\ref{fig5}. The magnetic field dependence of the phase shift is shown in the inset of Fig.~\ref{fig5}.   
Applying a gate voltage we found that for magnetic fields $B \gtrapprox 0.5$~T  the photovoltage exhibits $1/B$-periodic oscillations with amplitude substantially enhanced as compared to the response in the absence of an external magnetic field, see Figs.~\ref{fig6},~\ref{fig7}, and ~\ref{fig8}.  The period of the oscillations changes with variation of the gate voltage. In the vicinity of the CNP, the oscillations are almost absent, see Fig.~\ref{fig6}~(c). The change of the radiation frequency from 2.54 to 1.62~THz substantially enhances  the signal, but does not affect the oscillation period, see Fig.~\ref{fig7}. Comparison of the observed oscillations with the data on magnetotransport reveals that the oscillations follow the first derivative of the SdH magnetoresistance, see Fig.~\ref{fig8}. Increasing the radiation intensity by six order of magnitude, obtained by using high power pulsed laser, we observed that while edge photocurrents can clearly be detected, the magnetooscillations of the photoresponse vanish.

\section{Model}

In general, appearance of dc voltage in response to terahertz radiation can be caused by several phenomena including photogalvanic and photothermoelectric effects, as well as rectification of the terahertz induced dc electric current in gated structures and in a conducting channel supporting plasma waves~\cite{Dyakonov1996,Bandurin2018}. However, apart photogalvanics, all these mechanisms can hardly explain the key features of the observed edge dc current: (i) the edge current is driven by the radiation with a spatially uniform intensity and (ii) it is characterized by the specific polarization dependence described above, changing its direction when the polarization is reversed. The observed edge current also differs from the photocurrents formed in the edge channels of 2D topological insulators~\cite{Dantscher2017}, since such channels are not formed in our conducting samples.  To describe the experimental data we now demonstrate the microscopic model developed of the edge photocurrent induced by the high-frequency electric field. The photocurrent originates from $P$-symmetry breaking at the edge and may be viewed as consisting of two contributions. 

One of the contributions can be interpreted in terms of dynamic accumulation of electric charge near the edge and the synchronized charge driving along the edge. The microscopic mechanism is the following. The electric field component $E_x$ perpendicular to the edge causes back and forth motion of carriers in the sample giving rise to a charge depletion and enrichment at the sample edge. The $y$ component of the electric field drives the carriers along the edge at the same frequency. The interplay of both effects results in the emergence of a net $dc$ current flowing along the edge within the channel determined by the screening length of high-frequency field. 

\begin{figure}
	\centering
	\includegraphics[width=\linewidth]{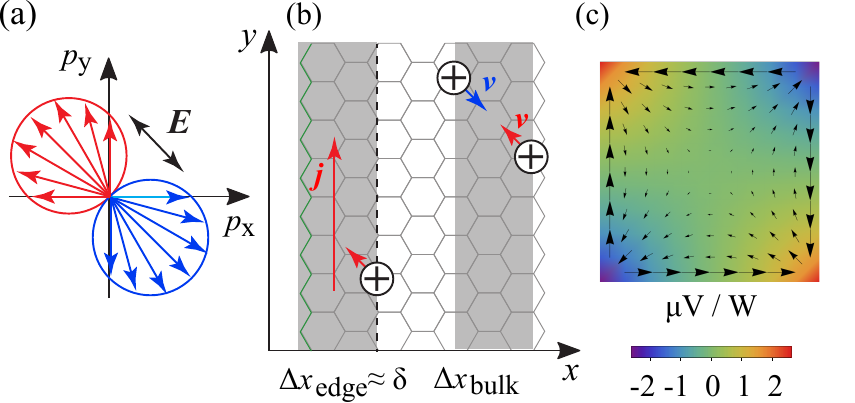}
	\caption{Microscopic model of the edge photocurrent formation sketched for $p$-type samples. (a) Optical  alignment of carrier momenta induced by the Drude absorption of linearly polarized terahertz radiation. Blue and red arrows illustrate the anisotropy in the distribution of carrier momenta ${\bm p}$ and, consequently, the velocities ${\bm v}$. (b) The carrier fluxes induced by the optical alignment in stripes of the mean free path width at the sample edge ($\Delta x_{\rm{edge}}$) and in the sample bulk ($\Delta x_{\rm{bulk}}$). Vertical red arrow shows the generated electric current caused by the imbalance of the fluxes at the edge. (c) Calculated distribution of the photocurrents (arrows) and photo-induced electrostatic potential normalized to the power of incident radiation in a square shaped sample for the azimuthal angle $\alpha = 45^\circ$. 
	}
	\label{model}
\end{figure}

The other contribution to the photocurrent stems from the alignment of the free carrier momenta by the high-frequency electric field. This mechanism of the current formation is illustrated in Fig.~\ref{model}. The Drude absorption of linearly polarized radiation leads to the alignment of the carrier quasi-momenta $\bm p$ where the carriers move predominantly along the line of the radiation electric field $\bm E$. The resulting distribution is sketched in Fig.~\ref{model}~(a) for the angle $\alpha = 45^\circ$ ~\footnote{Note that the phenomenon of optical alignment is know for inter-band and intra-band transitions in semiconductors and graphene. The specificity for the intra-band optical transitions is that the quasi-momenta get aligned along $\bm E$~\cite{Tarasenko2011} whereas for the inter-band transitions in graphene they are aligned orthogonally to $\bm E$~\cite{Golub2011,Hartmann2011}.}. The alignment results in an anisotropy of the carrier movement and is described by the second angular harmonic of the distribution function in the $\bm p$-space. It does not carry a net electric current. This is shown schematically for the holes in the bulk of BLG, see the right grey area indicated as $\Delta x_{\rm bulk}$ in Fig.~\ref{model}~(b). Here, two carrier fluxes coming from top left (blue arrow) and bottom right (red arrow) compensate each other and no electric current emerges. In the vicinity of the sample edge defined by the mean free path, $\Delta x_{\rm edge}$ in Fig.~\ref{model}~(b), the situation changes. Now, the flux from top left is absent. Consequently, uncompensated bottom right flux drives a net electric current $\bm j$ flowing along the edge. 

The model shown in Fig.~\ref{model}~(b) corresponds to hole conductivity. For electrons, the directions of the edge carrier fluxes remain the same, which means that the electric current is of opposite sign. The model also reveals that the direction of the edge current is defined by the orientation of the $ac$ radiation electric field with respect to the edge. The photocurrent vanishes for the radiation polarized along or normal to the edge and reaches a maximum at angles $\alpha = 45^\circ$ and $135^\circ$, having opposite signs for these angles. The overall polarization dependence is given by $j_{\rm edge} \propto \sin 2 \alpha$,  which corresponds to the experimental behavior of the photoresponse \footnote{A small phase shift $\varphi_0$ detected in the signal measured along $x-$axis and resulting in the difference of signal for $\alpha = 45^\circ$ and $135^\circ$ degrees, may be caused by the photocurrent excited in the bulk of bi-layer graphene. The mechanism of the bulk photocurrent is similar to that discussed for BiSbTe- based three dimensional topological insulators~\cite{Olbrich2014}, which belong to the same point group symmetry (C$_{3v}$) as bilayer graphene on a substrate as well as in the presence of an applied gate voltage and is out of scope of the present paper}.

For our experiments on a square-shaped sample with the laser spot larger than the sample, see Fig.~\ref{fig1}~(b), photocurrents emerge along all the edges and flow to the opposite corners of the sample. The continuity of the electric current is then provided by its spreading in the conducting bulk of the sample. The calculated distribution of the photocurrent and photo-induced electrostatic potential in the sample is illustrated in Fig.~\ref{model}~(c) for $\alpha = 45^\circ$. The details of the calculations are given in Sec.~\ref{discuss}.
All the features of the edge photocurrents discussed above, including the correlations in the current directions along the neighbouring edges of the sample and the polarization dependence, are indeed observed in the experiments, see Figs.~\ref{fig3} and \ref{fig4}. This unambiguously proves the microscopic origin of the observed photoresponse.

Classical magnetic field applied normally to the BLG plane results in the emergence of the Lorentz force acting upon moving carriers. This leads to a rotation of the electron distribution in the momentum space and, consequently, to a phase shift $\theta_B$ in the polarization dependence, which was observed in the experiments, see Fig.~\ref{fig5}. In quantizing magnetic fields, resulting in a formation of the Landau levels and the Shubnikov-de Haas oscillations in the conductivity. The experiment shows that the photocurrent also exhibits magnetooscillations with the same periodicity in $1/B$.

\section{Kinetic theory} \label{theory}

Now we develop a microscopic theory of the edge photogalvanic effect based on the Boltzmann kinetic theory~\cite{Karch2011}. In this approach, the distribution function of carriers $f(\bm p,x,t)$ 
is found from the equation
\begin{equation}\label{Boltzmann}
\pderiv{f}{t} + v_x \pderiv{f}{x} + e \left( \bm{{\cal E}}(x,t) + \frac{1}{c} \bm v \times \bm B \right) \pderiv{f}{\bm p}  = {\rm St} f \,,
\end{equation}
where $\bm p$ and $\bm v = \bm p /m^*$ are the momentum and velocity, respectively, $m^*$ is the effective mass, $e$ is the carrier electric charge, $\bm{{\cal E}}(x,t) = \bm{{\cal E}}(x) \exp(-\i \omega t) + {\rm c.c.}$ is the total electric field consisting of the $ac$ field of the irradiating wave $\bm{E} \exp(-\i \omega t) + {\rm c.c.}$ and the local $ac$ field with the amplitude $\delta E_x(x) \propto E$ induced by dynamical charge redistribution near the edge (see Appendix for details), $\bm B$ is the magnetic field, and ${\rm St} f$ is the collision integral. 
  
We solve Eq.~\eqref{Boltzmann} by expanding the distribution function in a series in the electric field amplitude as follows
\begin{equation}\label{f_exp}
f(\bm p, x ,t) = f_0 + [f_1(\bm p, x) \exp(-\i \omega t) + {\rm c.c.}] + f_2(\bm p, x) \,,
\end{equation}
where $f_0$ is the equilibrium distribution function, $f_1\propto E$ is the first-order correction, and $f_2 \propto E E^*$ is the time-independent second-order correction which determines the $dc$ current. Equations for $f_1$ and $f_2$ have the form
\begin{align}
-\i \omega f_1 + v_x \pderiv{f_1}{x} + e \bm{{\cal E}}(x)  \pderiv{f_0}{\bm p} + 
 \frac{e}{c} (\bm v \times \bm B)  \pderiv{f_1}{\bm p}  = {\rm St} f_1  \,, \label {f_1_equation} \\
v_x \pderiv{f_2}{x} + \left[ e \bm{{\cal E}}(x) \pderiv{f_1^*}{\bm p} + {\rm c.c} \right] + 
 \frac{e}{c} (\bm v \times \bm B)  \pderiv{f_2}{\bm p}  = {\rm St} f_2 \,.  \label{f_2_equation}
\end{align}

The local density of the $dc$ electric current $\bar{j_y}(x)$ is given by
\begin{equation}\label{j_y_def}
\bar{j_y}(x) = 4 e \sum \limits_{\bm p} v_y f_2(\bm p, x)  \,,
\end{equation} 
where the factor $4$ accounts for the spin and valley degeneracy. Multiplying Eq.~\eqref{f_2_equation} by $v_y$ and summing up the result over $\bm p$, we obtain
\begin{equation}\label{j_y_2}
\bar{j_y}(x) = - 4 e \tau_1 \sum \limits_{\bm p} v_x v_y \pderiv{f_2}{x} + 4 \frac{e^2 \tau_1}{m^*} \sum \limits_{\bm p} (E_y^* f_1 + E_y f_1^*) \,.
\end{equation}
Here, $\tau_1$ is the momentum relaxation time defined as $1/\tau_1 = - \sum_{\bm p} v_\alpha \, {\rm St} f / \sum_{\bm p} v_\alpha f$ and we took into account that the $dc$ current perpendicular to the edge is absent, i.e.,  $\sum_{\bm p} v_x f_2 = 0$.

The total electric current flowing along the edge is given by
\begin{equation}\label{J_y}
J_y =  \int\limits_{0}^{\infty} \bar{j_y}(x) dx \,.
\end{equation}
Using the relations $\sum_{\bm p} f_1 = - (i/\omega) \sum_{\bm p} v_x \partial f_1 /\partial x$, following from Eq.~\eqref{f_1_equation}, and $\sum_{\bm p} v_x f_1(\bm p,0) = 0$, which represents the lack of current through the edge, we obtain
\begin{align}\label{J_y_final}
J_y = &- 4 e \tau_1 \sum \limits_{\bm p} v_x v_y [f_2(\bm p, \infty) - f_2(\bm p, 0 )]  \\
&+ 4 i \frac{e^2 \tau_1}{\omega m^*} \sum \limits_{\bm p} [E_y v_x f_1^*(\bm p, \infty) - E_y^* v_x f_1(\bm p, \infty)] \nonumber \,.
\end{align}
Equation~\eqref{J_y_final} is quite general. It shows that the edge photocurrent consists of two contributions. The first one can be interpreted in terms of the alignment of the carrier momenta by the high-frequency electric field. The second contribution can be related to the dynamical charge redistribution near the edge (see Appendix). 

To proceed further, one needs to specify the boundary condition at $x=0$. We assume specular reflection of carriers at the sample edge implying that the distribution function satisfies $ f(p_x, p_y, 0) = f(-p_x, p_y, 0)$. In this case, the term $\sum_{\bm p} v_x v_y f_2 (\bm p, 0)$ vanishes and the edge current $J_y$ is determined by the corrections to the distribution function far from the edge where the $ac$ electric field is undisturbed, i.e., $\delta E_x(\infty) = 0$.

The  terms $\sum_{\bm p} v_x v_y f_2 (\bm p, \infty)$ and $\sum_{\bm p} v_x f_1 (\bm p, \infty)$ can be expressed via the tensor of bulk conductivity. Indeed, in the sample bulk, the amplitude of the $ac$ electric current induced by the $ac$ electric field is given by
\begin{equation}\label{sum_v_f_1}
j_{\alpha}^{({\rm b})}= 4e \sum_{\bm p} v_{\alpha} f_1(\bm p, \infty) = \sum_{\beta} \sigma_{\alpha \beta} E_{\beta} \,,
\end{equation}
where $\sigma_{\alpha \beta}$ is the tensor of linear conductivity,
\beqa
\label{sigma}
\sigma_{xx} &=& \sigma_{yy} = \frac{(1 - i \omega \tau_1) \sigma_0}{(1-i \omega \tau_1)^2 + (\omega_c \tau_1)^2}  \,, \\
\sigma_{xy} &=& - \sigma_{yx} = \frac{\omega_c \tau_1\, \sigma_0}{(1-i \omega \tau_1)^2 + (\omega_c \tau_1)^2} \,,  \nonumber
\eeqa
$\omega_c = e B_z/(m^* c)$ is the cyclotron frequency, $\sigma_0 = n e^2 \tau_1/m^*$, and $n$ is the carrier density. The expressions above can be readily obtained from Eq.~\eqref{f_1_equation}. In turn, Eq.~\eqref{f_2_equation} yields 
\beqa
&\sum \limits_{\bm p} v_x v_y f_2(\bm p, \infty) = \dfrac{\tau_2}{4 m^*} \label{sum_vv_f_2} \\
& \times \dfrac{ j_y^{(\rm b)} E_{x}^* + j_x^{(\rm b)} E_{y}^* - 2 \omega_c \tau_2 ( j_x^{(\rm b)} E_{x}^* - j_y^{(\rm b)} E_{y}^*)}
{1+(2 \omega_c \tau_2)^2}  + {\rm c.c.} \nonumber \,,
\eeqa
where $\tau_2$ is the relaxation time of the second angular harmonic, $1/\tau_2 = - \sum_{\bm p} v_x v_y \, {\rm St} f / \sum_{\bm p} v_x v_y f$. 

Finally, taking into account Eqs.~\eqref{J_y_final},~\eqref{sum_v_f_1}, and~\eqref{sum_vv_f_2}, we obtain the edge photocurrent
\beqa\label{J_y_final2}
& J_y  = \dfrac{e\tau_1}{m^* \omega} \left[{\rm Im} \, \sigma_{xy} \, |\bm E|^2  - {\rm Re}\, \sigma_{xx}\, \i (E_x E_y^* - E_y E_x^*)  \right]  \;\;\; \\
& -  \dfrac{e\tau_1 \tau_2}{m^*} \left[ \dfrac{2{\rm Re}(\sigma_{xx} - 2 \omega_c \tau_2 \sigma_{xy})}{1+(2\omega_c \tau_2)^2} 
- \dfrac{{\rm Im} \sigma_{xx}}{\omega \tau_2} \right] (E_{x}E_{y}^* + E_{y}E_{x}^*) \nonumber \\
& +  \dfrac{e\tau_1 \tau_2}{m^*} \left[ \dfrac{2{\rm Re}(\sigma_{xy} + 2 \omega_c \tau_2 \sigma_{xx})}{1+(2\omega_c \tau_2)^2} 
- \dfrac{{\rm Im} \sigma_{xy}}{\omega \tau_2} \right] (|E_{x}|^2 - |E_{y}|^2) \,. \nonumber 
\eeqa

The edge photocurrent~\eqref{J_y_final2} contains (i) the polarization-independent term $J_y \propto |\bm E|^2$, (ii) the contribution sensitive to the circular polarization of incident radiation $J_y \propto \i (E_x E_y^* - E_y E_x^*)$, and (iii) the terms proportional to the Stokes components of the incident radiation $E_{x}E_{y}^* + E_{y}E_{x}^*$ and $|E_{x}|^2 - |E_{y}|^2$ which define the linear polarization. Note that, at zero magnetic field, the terms $\propto |\bm E|^2$ and $\propto (|E_{x}|^2 - |E_{y}|^2)$ vanish and the polarization dependence of the photocurrent excited by linearly polarized radiation is given by $E_{x}E_{y}^* + E_{y}E_{x}^*$. 

At $\omega \gg \omega_c$, relevant to our experimental conditions, and $\tau \equiv \tau_1 = \tau_2$, the photocurrent~\eqref{J_y_final2} takes the form
\begin{multline}\label{J_y_final3}
J_y =  \frac{n e^3 \tau^3}{m^{*2} (1 + \omega^2 \tau^2)}  \left[ \frac{2 \omega_c \tau |\bm E|^2}{1 + \omega^2 \tau^2}  - \frac{\i (E_x E_y^* - E_y E_x^*)}{\omega \tau}    \right. \\
\left. - \frac{E_{x}E_{y}^* + E_{y}E_{x}^*}{1 + 4 \omega_c^2 \tau^2} + \frac{2 \omega_c \tau (2 + \omega^2 \tau^2) (|E_{x}|^2 - |E_{y}|^2)}{(1 + 4 \omega_c^2 \tau^2) (1 + \omega^2 \tau^2)}  \right] \:.
\end{multline}
In particular, at $\omega \tau \gg 1$, the dominant contribution to the photocurrent is given by the last two terms. In this case, the effect of magnetic field is reduced to a decrease in the photocurrent magnitude and a shift in the polarization dependence determined by the angle $\theta_B = \arctan (2 \omega_c \tau)$.

\section{Discussion} \label{discuss}

The above kinetic theory of the edge photogalvanic effect is developed for classical magnetic fields. The conditions relevant to our experiment are $\omega \tau \gg 1$ and $\omega \gg \omega_c$.
The last two terms $J_y \propto (E_{x}E_{y}^* + E_{y}E_{x}^*)$ and $J_y \propto (|E_{x}|^2 - |E_{y}|^2)$ in Eq.~\eqref{J_y_final3}  describe the observed polarization dependence of the photocurrents picked up along the adjacent edges. At zero magnetic field the edge photocurrent $J_y \propto (E_{x}E_{y}^* + E_{y}E_{x}^*) \propto \sin 2 \alpha$ (Fig.~\ref{fig3}), whereas in classical magnetic fields its polarization dependence gains a phase shift (Fig.~\ref{fig5}). As follows from Eq.~\eqref{J_y_final3} at $\omega \tau \gg 1$, the photocurrent, and correspondingly the photovoltage, $U_y \propto J_y \propto \sin (2\alpha + \theta_B)$, where the magnetic field dependence of the phase shift is given by $\theta_B = \arctan(2\omega_c \tau)$. This behavior agrees well with that observed in the experiment (inset of Fig.~\ref{fig5}) for $m^* = 0.03~\mathrm{m_0}$ and $\tau = 0.6$~ps. Note, that this value of $\tau$ is in agreement with an estimation $\tau \sim 0.2$~ps based on typical values of mobility $\mu \sim 10^4$~cm$^2$/Vs in our samples. Equation~\eqref{J_y_final3} also explains the opposite signs of the photocurrents for $p$- and $n$-type conductivities (Figs.~\ref{fig3} and~\ref{fig4}): Since $J_y \propto e^3$, the currents directions are opposite for electrons ($e<0$) and holes ($e>0$).

Furthermore, Eq.~\eqref{J_y_final3} yields the correct magnitude of the photoresponse. In the experiments, we measure photovoltage between contacts in the open circuit configuration. To calculate the spacial distribution of the photo-induced electrostatic potential $\Phi(x,y)$ and current spreading in the sample we solve the continuity equation for the total $dc$ current $\nabla \cdot (\bar{\bm j} + \bm j^{\rm dr}) = 0$, where $\bar{\bm j}$ is the edge photocurrent and $ \bm j^{\rm dr}$ is the compensating drift current flowing in the bulk of the sample. Taking into account that $j^{\rm dr}_\alpha = \sum_\beta \sigma_{\alpha \beta} \nabla_\beta \Phi$, where $\sigma_{\alpha \beta}$ is the $dc$ conductivity given by Eq.~\eqref{sigma} at $\omega = 0$, we obtain the Poisson equation
\begin{equation}
\label{Phi}
\frac{\sigma_0}{1 + \omega_c^2 \tau_1^2} \Delta \Phi = \nabla \cdot \bar{\bm j}\:.
\end{equation}
This equation complemented with the boundary condition of zero electric current $\bar{\bm j}_{\bm n} + \bm j^{\rm dr}_{\bm n}$ across the sample edges is solved numerically using Green's function method as in Ref.~\cite{Budkin2016}. The spatial distribution $\Phi(x,y)$ calculated from Eqs.~\eqref{J_y_final3} and~\eqref{Phi} and the corresponding total $dc$ current are shown in Fig.~\ref{model}~(c) for $\omega \tau \gg 1$, $\alpha = 45^\circ$, zero magnetic field, the effective mass $m^* = 0.03~\mathrm{m_0}$, the radiation frequency $\omega/(2\pi) = 2.54$ THz, and the laser spot diameter $1.5$ mm much larger than the sample size. The calculated photovoltage between the neighbouring corners of the sample is $V/P \approx 4$~$\mu$V/W, which agrees well with the measured photovoltage amplitude and sign (Fig.~\ref{fig3}) as well as the simple analytical estimation $V \sim J_y/\sigma_0 \sim e E^2 /(m^* \omega^2)$. Note that the measured photocurrent amplitude  is by about an order of magnitude larger than that of the  edge photocurrent previously reported for the single layer graphene~\cite{Karch2011}.

In strong magnetic fields, the density of states $D(\varepsilon)$, and accordingly the relaxation time $\tau$, acquires oscillating dependence on the electron energy originating from the formation of Landau levels. This will lead to additional terms in the edge photocurrent $J_y \propto d D(\varepsilon) / d \varepsilon$, not considered in the above theory, which may dominate and determine the oscillating dependence of the photocurrent on magnetic field.  At high electron temperatures, the SdH-oscillations, and related magnetooscillations of the photocurrent, vanish, as detected in the experiments applying high power radiation of the pulsed THz laser. Note that, under these conditions, photothermoelectric effects may contribute to the signal. Development of a microscopic theory of the edge photogalvanic effect in the regime of the Shubnikov-de Haas oscillations is a task for the future.

\section{Summary} 
\label{summary}

Combining the experimental data and theory we have shown that the optical alignment of the free carrier momenta and dynamic charge accumulation at the edges of the bi-layer graphene samples, caused by linearly polarized terahertz field, drive a direct electric current. The photocurrent is formed within the channels at the sample edges, whose width is defined by the mean free path and the screening length of the terahertz field. The observed features of the edge photocurrent excited at zero and classical magnetic fields as well as the current magnitude are well described by the developed theory. In quantizing magnetic fields, the photocurrent exhibits sign-alternating magnetooscillations which are periodic in $1/B$, similarly to the Shubnikov-de Haas oscillations in conductivity. Even stronger magnetic fields (not achieved here) would realize the quantum Hall effect regime with the  topological chiral edge channels responsible for the conductivity and photoresponse~\cite{Plank2019}. Our results suggest that second-order nonlinear processes can be quite efficient in devices of mean free path size.

\section{Acknowledgments}
\label{acknow}
We thank G.V. Budkin for fruitful discussions.
The support from the Deutsche Forschungsgemeinschaft (DFG, German Research Foundation) -  Project-ID 314695032 - SFB 1277, the Elite Network of Bavaria (K-NW-2013-247), the Volkswagen Stiftung Program (97738), and the RFBR (grants 19-02-00095 and 19-02-00825) is gratefully acknowledged.  A.M. acknowledges the support of EPSRC Early Career Fellowship EP/N007131/1. M.V.D. acknowledges financial support from the Russian Science Foundation (project 19-72-00029).  VF acknowledges support from European Graphene Flagship Project, ERC Synergy Grant Hetero2D, EPSRC grants EP/S030719/1 and EP/N010345/1.

\appendix
 
\section*{Appendix: Dynamic charge accumulation at the edge}

The contribution to the $dc$ edge current caused by dynamical modulation of the carrier density near the edge is given by
\beq \label{Jydrift}
\delta J_y = \int \frac{\d \sigma_0}{\d n}  \left[ \delta n(\bm{{\cal E}}, x) E^*_y + \delta n^*(\bm{{\cal E}},x) E_y \right] dx  \,, 
\eeq
where $\delta n(\bm{{\cal E}},x) = 4 \sum_{\bm p} f_1$ is the local correction to the carrier density linear in the $ac$ electric field $\bm E$ and $\sigma_0 = n e^2 \tau/m^*$, see the second term in the right-hand side of Eq.~\eqref{j_y_2} and Ref.~\cite{GlazovGanichev_review}. 

The profiles of the charge accumulation $e \, \delta n(\bm{{\cal E}}, x)$ and the total electric field $\bm{{\cal E}} = \bm{E} + \delta \bm E(x)$   should be calculated self-consistently. We find the profiles by solving the continuity equation for the electron density and current
\beq
\d_x j_x = -i\, e  \,\omega \, \delta n \,,
\eeq
together with the ``drift and diffusion equation'' 
\beqa \label{drift1}
 j_x &=& e\, D_{xx} \,\d_x \delta n + \sigma_{xy} E_y +\\ \nn && 
 +\sigma_{xx} \left[ E_x + \frac{2 \pi e}{\eps r}\int_0^\infty {\mathcal K}\left(\frac{x-x'}{r}\right) \delta n(x') dx'\right] \,.
\eeqa   
Here, $D_{xx}$ is the diffusion coefficient, the electric field along $x$ is the sum of the driving external field $E_x$ and the field $\delta E_x$ due to the dynamically induced charge near the edge~\cite{Volkov88},

${\mathcal K}$ is the Coulomb interaction kernel
$$
   {\mathcal K}(\xi) = -\frac{2 \mbox{Ci}\left(|\xi|\right) \sin \left(|\xi|\right)+\left(\pi -2 \mbox{Si}\left(|\xi|\right)\right) \cos
   \left(|\xi|\right)}{2 \pi} \sign(\xi),
$$
which describes the electric field of a charged wire, and it is assumed that $\omega \tau <1$.  The expression for  $\delta E$ is derived from the 1D Fourier transform of the Keldysh potential~\cite{Keldysh1979}
\beq \nn
V(q) = \frac{2 \pi e^2}{\eps q (1+ r q)} \,,
\eeq
where $\eps$ is the effective dielectric constant of the substrate, $\eps = \left(1+\eps^{\rm SiO_2}\right)/2$, and the thickness 
$$r = d^{\rm hBN} \frac{\eps^{\rm hBN}-1}{2 \eps}$$
takes into account the in-plane polarizability of the encapsulating hBN film of the total thickness $d^{\rm hBN}$ and its dielectric constant $\eps^{\rm hBN}$. 

Differentiating both sides of Eq.~\eqref{drift1} with respect to $x$ and using the continuity equation, we obtain
\beqa
i \, \omega \, \delta n = -D_{xx} \d_x^2 \delta n -  
\frac{2 \pi \sigma_{xx}}{\eps r^2} \int_0^\infty \d{\mathcal K}\left(\frac{x-x'}{r}\right) \delta n(x') dx'  \nn , \\
\d {\mathcal K}(\xi) \equiv \frac{[\pi -2 \mbox{Si}(|\xi|)] \sin(|\xi|)-2 \mbox{Ci}(|\xi|) \cos(\xi)}{2 \pi} - \delta(\xi) . \nn
\eeqa

The boundary conditions, $j_x(0)=0$ and $j_x(\infty) =  \sigma_{xx}  E_x + \sigma_{xy} E_y$, lead to a normalization condition for the carrier accumulation $\delta n(x)$,
\beq
\Delta N \equiv \int_0^\infty \delta n(x) dx = \frac{\sigma_{xx}  E_x + \sigma_{xy} E_y}{i \omega e }.
\eeq
Note, that the total carrier accumulation amplitude $\Delta N$  is independent of the Coulomb interaction strength. According to Eq.~\eqref{Jydrift}, the screening of the electric field at the edge does not affect $\delta J_y$ either 
since the field amplitude $E_y$ is constant and the integral $\int \delta n(x) E_y^* dx = \Delta N E_y^*$ is independent of Coulomb interaction.

However, the Coulomb interaction changes the spatial profile of the edge current.  To proceed, we introduce the length $l_\mathrm{eff} = \sqrt{2 D_{xx}/\omega}$
and the dimensionless  parameters  
\beq
\tilde r = r/l_\mathrm{eff} \ , \ \ \kappa = \frac{2 \pi \sigma_{xx}}{\eps\, \omega\, r } 
\eeq
and present the ``drift and diffusion equation'' in the form
\beq \label{profile}
i\, \delta n = -\frac12 \d_{\tilde x}^2 \delta n +  
\frac{\kappa}{\tilde r} \int_0^\infty \d {\mathcal K}\left(\frac{\tilde x - \tilde x'}{\tilde r} \right) \delta n(\tilde x') d \tilde x'. 
\eeq

\begin{figure}
  \includegraphics[width= \columnwidth]{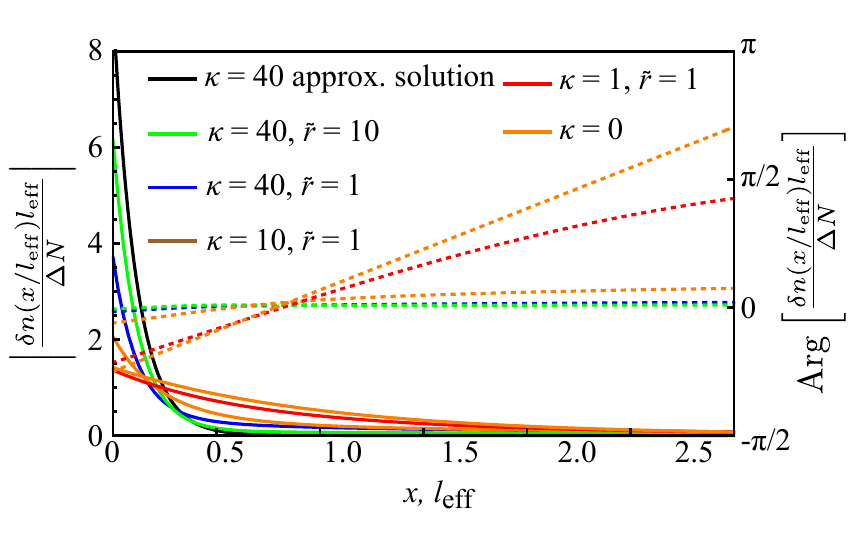}
\caption{ \label{fig:Plots} 
The spatial profiles of the magnitude $|\delta n / \Delta N|$ (solid lines) and the phase $\mathrm{Arg} (\delta n / \Delta N)$ (dashed lines) of charge accumulation.}  
\end{figure}

\begin{figure}
  \includegraphics[width= \columnwidth]{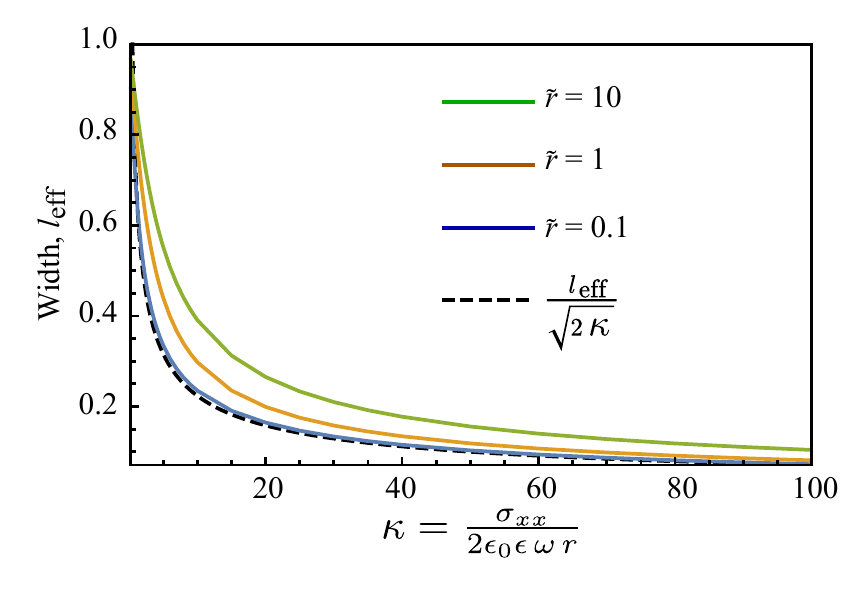}
\caption{ \label{fig:StripeWidth} 
Width of carrier accumulation stripe $w$ as a function of $\kappa$ calculated for several values of $\tilde r$. Dashed curve shows the analytical asymptotics plotted after Eqs.~\eqref{approxprofile} and~\eqref{asympt}.}  
\end{figure}

Numerical solutions of Eq.~\eqref{profile} are illustrated in Fig.~\ref{fig:Plots} and show exponentially decaying profiles $\delta n(x)$ 
\beq \label{approxprofile}
\delta n(x) \approx k \frac{\sigma_{xx} E_x + \sigma_{xy} E_y}{i \omega\, e \,l_\mathrm{eff}} e^{-k x/l_\mathrm{eff}},
\eeq
where $k$ is found from the equation
\beq \label{Eq_k}
i = -k^2/2 + \kappa  \frac{\left(2 \pi  k^2 \tilde r^2-2 k \tilde r \log (k \tilde r)+\pi \right)}{2 \pi (k^2 \tilde r^2+1)} \,.
\eeq
The solution of Eq.~\eqref{Eq_k} for real $\kappa$ in the limiting cases of small and large $\kappa$ has the form
\beq \label{asympt}
  k = \left\{\ba{cc}1-i, & \kappa\ll 1 , \\  
  \sqrt{2 \kappa} 
  , & \kappa \gg 1 \ea \right. .
\eeq
%
The dependence of the width of carrier accumulation stripe $w$ defined by $|\delta n(w)/ \delta n(0)|= 1/\rm e$ on $\kappa$ is plotted in Fig.~\ref{fig:StripeWidth} for several values of $\tilde r$. Dashed curve shows the analytical asymptotic plotted after Eqs.~\eqref{approxprofile} and~\eqref{asympt}. Overall,  the width of carrier accumulation near the edge interpolates from $l_\mathrm{eff} = \sqrt{2 D_{xx}/\omega}$ to $l_\mathrm{eff}/\sqrt{2 \kappa} = \sqrt{D_{xx} \eps r / (2 \pi \sigma_{xx})} = (\hbar/2) \sqrt{\eps r/(e^2 m_*)}$ with increasing $\kappa$, see Fig.~\ref{fig:StripeWidth}.  Having estimated the feasible value of $\kappa$, we find that $| \kappa | \gg 1$ and the width of the charge accumulation stripe is much smaller than $l_\mathrm{eff}$. The Coulomb contribution to the electric field $\delta E_x(x) = 2\pi e /(\eps r) \int_0^\infty {\mathcal K}[(x-x')/r] \delta n(x')\, dx' $ increases the total electric field right near the edge and decreases it away from the edge in such a way that $\int_0^\infty \delta n(x)\, \delta E_x(x)\, dx = 0$ which follows from the antisymmetry of ${\mathcal K}$.  

\bibliography{edgecurrents}

\end{document}